\documentclass[twocolumn,showpacs,preprintnumbers,amsmath,amssymb,prl]{revtex4}
\usepackage{amsbsy}
\usepackage{graphicx,color}
\usepackage{amsfonts}
\usepackage{amsmath}
\usepackage{amssymb}
\begin{document}
\title{"Gray" BCS condensate of excitons and internal Josephson effect}
\author{R. Combescot$^{(a),(b)}$ and M. Combescot$^{(c)}$}
\address{(a) Laboratoire de Physique Statistique, Ecole Normale Sup\'erieure, UPMC  
Paris 06, Universit\'e Paris Diderot, CNRS, 24 rue Lhomond, 75005 Paris,  
France}
\address{(b) Institut Universitaire de France,103 boulevard Saint-Michel, 75005 Paris, France}
\address{(c) Institut des Nanosciences de Paris, UPMC Paris 06, CNRS, 2 pl. Jussieu, 75005 Paris, France}
\date{Received \today}

\begin{abstract}
It has been recently suggested that the Bose-Einstein condensate formed by excitons in the dilute limit must be dark, i.e., not coupled to photons. 
Here, we show that, under a density increase, the dark exciton condensate must acquire a bright component due to carrier exchange in which dark excitons turn bright. This however requires a density larger than a threshold which seems to fall in the forbidden region of the phase separation
between a dilute exciton gas and a dense electron-hole plasma. The BCS-like condensation which is likely to take place on the dense side, must then have a dark and a bright component - which makes it "gray". It should be possible to induce an internal Josephson effect between these two coherent
components, with oscillations of the photoluminescence as a strong proof of the existence for this "gray" BCS-like exciton condensate.
\end{abstract}
\maketitle

Wannier excitons have been under extensive studies for decades. Made of
two fermions, they are bosonic particles. So, as pointed out long ago \cite{bbb},
a dilute gas of excitons should undergo a Bose-Einstein condensation. 
The carrier mass being very light and
Coulomb attraction quite reduced by the large crystal dielectric constant, the exciton Bohr radius $a_X$ is two orders of
magnitude larger than the hydrogen atom Bohr radius. So, many-body effects controlled by the dimensionless parameter
$\eta=N(a_X/L)^D$ where $N$ is the exciton number, $L$ the sample size and $D$ the space dimension, can easily be made
significant for $N$ not too large. The most dramatic one is the Mott dissociation of excitons into an electron-hole plasma \cite{shah} when the
distance between two excitons is of the order of their size. Actually, $\eta \sim 1$ most often fall in an instability region with phase separation between a dilute exciton phase and a dense electron-hole phase. In Si and Ge
\cite{comnoz,BRA}, this exciton dissociation is spontaneous at $T=0$ because the lowest energy phase is the high density electron-hole plasma
which is stabilized by the multivalley structure of the conduction band. In the case of direct gap semiconductors, a similar phase separation
can occur at $T=0$, but only under a density increase \cite{MC,loz}.

The exciton composite nature also appears through the fact that excitons come in
"bright" and "dark" states. Bright excitons, coupled to $\sigma_{\pm}$ photons, are made of $(\mp 1/2)$
conduction electrons and $(\pm 3/2)$ valence holes. Carrier exchanges however transform two opposite spin bright excitons (+1) and (-1) into two
dark excitons (+2) and (-2), these ($\pm 2$) excitons being made of ($\pm 1/2$) conduction electrons and ($\pm 3/2$) valence holes.
These dark excitons have actually a lower energy than bright excitons. Indeed, in addition to the intraband Coulomb processes
responsible for the dominant part of the exciton binding energy, interband Coulomb processes also exist in which one conduction
electron returns to the valence band while one valence electron jumps in a conduction state. By just writing \cite{Leunberger} that the electron conserves
its spin when it changes from a conduction state with orbital momentum $l=0$ to a valence state with orbital momentum $l=1$, it is possible to
show that the electron-hole pair which undergoes these interband processes must be bright. 
Since Coulomb interaction between electrons is repulsive, bright excitons thus
have an energy slightly higher than the dark exciton energy. As a result, if a Bose-Einstein condensation of excitons occurs,
this must be in these lowest energy dark states \cite{cbc}. Note that, even if excitons are formed from photon absorption in states which
are bright by construction, two opposite spin bright excitons can turn dark by carrier exchange.

Experimentally, the search for an exciton Bose-Einstein condensate has essentially
been made up to now through photoluminescence experiments. With  a dark exciton condensate, there was definitely no hope to evidence it in this way.
To possibly "see" a dark condensate, it is somewhat mandatory to first know where it is located spatially, i.e., to trap it. The center of the trap should then turn
dark when condensation occurs \cite{Leunberger}. Snoke and coworkers \cite{snoke} have seen such a darkening under a temperature decrease, using a trap obtained by applying
a local negative pressure. A cleaner way to trap excitons has been recently proposed \cite{piermaro} through two counter-propagating 
laser beams with linear polarization. The field modulation can then equally trap bright and dark excitons through carrier exchanges 
with different components of the light.

However, there is a major common problem in all experimental searches for exciton Bose-Einstein condensation.
Since experiments are performed at some not so low temperature, one needs an exciton density $n=N/L^D$ high enough
to have the critical temperature $T_c$ for the condensation above the experimental temperature, since for example, in 3$D$,
$T_c \sim n^{2/3}$ in the case of dilute gases. On the other hand, the fact that the dark exciton condensation prevails over
condensation in bright states is a result which has been obtained in the dilute limit, where interactions can be safely ignored.
One may wonder if this remains valid when the exciton density is high enough for interactions to become significant.

In this paper, we will show that, under a density increase, the dark exciton condensate must acquire a bright component, 
due to carrier exchange in which dark excitons turn bright. We find that this occurs only when the density is larger than some
threshold, the existence of a fully dark condensate \cite{cbc} being valid over a whole range of low densities. Nevertheless this
appearance of a bright component is in itself quite interesting. This makes the condensate look
"gray", which should make it much easier to observe. However, our evaluation of the density threshold for standard situations
puts the appearance of this gray condensate in the forbidden region of the Mott dissociation transition. Hence, this  gray condensate should
actually appear in a dense electron-hole plasma, and no longer in a fairly dilute exciton gas. However, it is not possible to have strictly speaking
a Bose-Einstein condensation in a dense electron-hole system. This would make this gray condensate inobservable.

Nevertheless, a similar condensation may occur in a dense system, as pointed out by Keldysh and Kopaev
\cite{kk}. Instead of an excited semiconductor, they considered a semimetal with both electrons and holes present at thermodynamical
equilibrium. They showed that the Coulomb attraction between an electron and a hole, although strongly reduced by free carrier screening, 
leads to the formation of pairs, in the same way as the small phonon-mediated attraction between two electrons produces 
a BCS condensate of Cooper pairs 
in standard superconductivity. Such a "Cooper pair" would then be a bound state of a
positively charged electron and a negatively charged hole, in the presence of an electron Fermi sea and of a hole Fermi sea. 
This pair, quite analogous to an exciton, might be called "excitonic Cooper pair". 
A similar condensation is expected to occur in a photocreated electron-hole plasma, turning it into a superfluid condensate.

Actually, this superfluid of excitonic Cooper pairs is not drastically different from the dark BEC condensate we have considered. 
Indeed, early works by Eagles \cite{eagles} in the sixties and more recent works by
Leggett \cite{leg} and by Nozi\`eres and Schmitt-Rink \cite{nsr}, have shown that one 
can continuously go from a Bose-Einstein condensate to a BCS-like condensate. 
$^6$Li and $^{40}$K ultracold fermionic gases provide
remarkable experimental realizations of this BEC-BCS crossover \cite{gps}. 

An important difference with atomic gases however is that no liquid-gas first order transition has been seen in these gases.
In contrast, such a phase separation is expected to occur in semiconductors \cite{MC,loz}. It should be stressed
that a very analogous situation exists in the BEC-BCS transition of deuterons in symmetrical nuclear matter \cite{urban}. In this case too, the
dilute side corresponds to a gas of deuterons whereas, on the dense side, one has Cooper-pair-like proton-neutron correlations in the presence
of Fermi seas. In this nuclear case, a liquid-gas phase transition is known to occur, in contrast with cold atomic gases.

In order to establish the appearance of this bright component on a strong basis, we are going to concentrate on the dilute side.
Contact with the possible existence of a "gray" phase in the dense electron-hole plasma phase will be made by extrapolating our result to higher densities.
Obviously, we can not securely claim that this extrapolation will be quantitatively valid, but we can reasonably expect it to be at least qualitatively correct.
Moreover, it is worth noting that the range of validity of such extrapolations are often much wider than what is a priori expected. Monte-Carlo calculations  \cite{gps} for example show that, in the case of fermionic ultracold gases, the equation of state based on the Lee-Huang-Yang expansion \cite{lhy} stays valid quite far toward the dense regime where interactions are very strong.

To present our idea in the simplest way, we first use an oversimplified model. A more realistic description of the problem will be given in a second step.
We first omit a part of the fourfold exciton degeneracy and only consider one kind of bright excitons with creation operators $b_{\bf k}^{\dag}$ and
one kind of dark exciton with creation
operators $d_{\bf k}^{\dag}$ with ${\bf k}$ being the exciton center-of-mass momentum. In the very dilute limit, the exciton effective Hamiltonian
reduces to its kinetic energy terms
\begin{eqnarray}\label{eq1}
H_{\rm kin}= \sum_{\bf k}\frac{k^2}{2m_{X}}\;d_{\bf k}^{\dag}d_{\bf k} + \sum_{\bf k}\left(\epsilon_0+\frac{k^2}{2m_{X}}\right)
\;b_{\bf k}^{\dag}b_{\bf k}
\end{eqnarray}
where $m_{X}$ is the exciton mass, and $\epsilon _0$ 
is the dark exciton binding energy compared to the bright state.
Since $\epsilon _0$ is positive, of order of a few tens of $\mu $eV, the $T=0$ ground state is a Bose-Einstein condensate of dark excitons.
However, this conclusion changes outside the very dilute regime due to
interactions. At low temperature, we may restrict them to their s-wave
component, which implies that we can take them wavevector independent. In our oversimplified model, we only keep the term
which is crucial to our conclusion, namely the one which describes the conversion of two dark excitons into two bright excitons and
conversely. The interaction term then reduces to
\begin{eqnarray}\label{eq2}
H_{\rm int}=g_{db} \sum_{{\bf k}_i}\,\left(b_{{\bf k}_1}^{\dag}b_{{\bf k}_2}^{\dag}d_{{\bf k}_3}d_{{\bf k}_4}+{\rm h.c.}\right)
\end{eqnarray}
with ${\bf k}_1+{\bf k}_2={\bf k}_3+{\bf k}_4$ due to momentum conservation.
For a macroscopic occupancy of dark exciton states, this term brings an induced macroscopic occupancy of bright states.
Let us now show that this is energetically favorable. Such an idea is somewhat analogous to what occurs in a two-level system where any coupling between levels lowers the ground state energy.

We can handle this problem to lowest order in the interaction by using mean field theory.
The macroscopic occupancies of the dark and bright exciton states with momentum ${\bf k}_i={\bf 0}$, then appear through the
standard  \cite{agd} mean field substitution $d_{\bf 0} \rightarrow \sqrt{N_d}\,e^{i\varphi_d}$ and $b_{\bf 0} \rightarrow \sqrt{N_b}\,e^{i\varphi_b}$,
where $N_d$ and $N_b$ are the number of dark and bright excitons in the sample,  $\{N_d,\varphi_d\}$ being conjugate variables, as well as $\{N_b,\varphi_b\}$.
The Hamiltonian $H=H_{\rm kin}+H_{\rm int}$ then has the classical limit:
\begin{eqnarray}\label{eqh}
{\mathcal H}=\epsilon _0 N_b+2g_{db}N_d N_b \cos\left[2(\varphi_b - \varphi_d)\right]
\end{eqnarray}
We note, from dimensional arguments, that $g_{db}$ must depend on sample size $L$ as $1/L^D$.

The above expression is very similar to the classical Hamiltonian appearing, for example, 
in a Josephson junction \cite{ajlbook}, in the macroscopic description 
of atomic Bose-Einstein condensate in double wells \cite{cctdgo} or in the dynamics \cite{legHe} of superfluid $^3$He phases.
Analogy with this last case leads us to predict the existence of an internal Josephson effect in the excitonic condensate,
with an internal flow of excitons between the dark and the bright condensate, associated to the oscillation of the relative phase
$\varphi \equiv (\varphi_b - \varphi_d)$ of these two components. Since the magnetic moments of dark and bright
excitons are different, it should be possible to excite this Josephson oscillation by a r.f. field.

Coming back to Eq.(\ref{eqh}), we see that the energy is minimum for $2\varphi=0$ or $\pi $,
depending on the sign of $g_{db}$; so, $g_{db}\cos (2 \varphi )$
can be replaced by $-|g_{db}|$  at the minimum. 
Minimization of the total energy at fixed exciton number $N=N_d+N_b$ then leads to a bright exciton number, at the minimum, equal to:
\begin{eqnarray}\label{eqnb}
N^{(0)}_b=\frac{N}{2}-\frac{\epsilon _0}{4|g_{db}|} \equiv \frac{N-N_c}{2}
\end{eqnarray}
A threshold $N_c = \epsilon _0/2|g_{db}|$ thus exists for the appearance of a bright component in the condensate. 
For $N < N_c$, the Bose-Einstein condensate is made of dark
excitons only, as in the very dilute limit, and there is no bright excitons at $T=0$. By contrast, for $N > N_c$, the condensate
is made of a coherent combination of a dark and a bright condensate, namely $(d^{\dag})^{N-N^{(0)}_b}(b^{\dag})^{N^{(0)}_b}|vac\rangle$
at zeroth order in the interaction.

To derive the equations ruling the internal Josephson effect, we proceed as usual. The total number of excitons $N=N_d+N_b$ being conserved, it is convenient to 
take $\delta N \equiv (N_b - N_d)/2$ and $\varphi$ as conjugate variables. The corresponding Hamilton equations 
$\dot{\delta N}=\partial {\mathcal H}/\partial \varphi$ and $\dot{\varphi}=-\partial {\mathcal H}/\partial \delta N$ then describe the Josephson
effects. In particular, at small departure from equilibrium Eq.(\ref{eqnb}), the resulting linear equations give rise to harmonic
oscillations with frequency $\omega _J$:
\begin{eqnarray}\label{}
\hbar^2\omega _J^2 = 32 g^2_{db}N^{(0)}_d N^{(0)}_b=2 \epsilon _0^2 \left(\frac{N^2}{N_c^2}-1\right)
\end{eqnarray}
For $\epsilon _0 \simeq 10 \mu {\rm eV}$, this gives $\omega _J \simeq 10^{10} N/N_c$, far from threshold, putting it in upper GHz range.
On the other hand this frequency goes to zero when $N=N_c$. Note that it should be possible to change $N_c$ by applying a magnetic field.

We now consider a more precise description of the problem, still on the dilute side. We take into account the fact that dark and bright excitons
come in two polarizations. Since the exciton wavevectors ${\bf k}$ are equal to zero in the condensed state, we can avoid writting them.
We note as $b_{\pm}^{\dag}$ the creation operator of bright exciton with spin $\pm 1$ and 
similarly $d_{\pm}^{\dag}$ for dark exciton with spin $\pm 2$. The interaction Hamiltonian, 
with all possible scattering processes between dark and bright excitons, then reads
\begin{eqnarray}\label{eq2a}
H_{\rm int}=v_{dd} \sum_{\sigma}\,d_{\sigma}^{\dag}d_{\sigma}^{\dag}d_{\sigma}d_{\sigma}
&+&v_{bb} \sum_{\sigma}\,b_{\sigma}^{\dag}b_{\sigma}^{\dag}b_{\sigma}b_{\sigma}  \\
+v_{db} \sum_{\sigma \sigma'}\,d_{\sigma}^{\dag}b_{\sigma'}^{\dag}b_{\sigma'}d_{\sigma}&+&g_{db} \sum_{\sigma}\,\left(b_{\sigma}^{\dag}b_{-\sigma}^{\dag}d_{-\sigma}d_{\sigma}+{\rm h.c.}\right)  \nonumber 
\end{eqnarray}
with $(\sigma, \sigma')=\pm$. The two first terms of $H_{\rm int}$ describe the effective scattering between two dark or 
two bright excitons with same spin. The third term describes the effective scattering between a bright and a dark exciton. The last term,
which is the conversion term considered in Eq.(\ref{eq2}), describes the scattering between two dark excitons with opposite spins into two bright excitons
with opposite spins. 

The mean field substitution now reads 
$d_{\pm} \rightarrow \sqrt{N_{\pm 2}}\,e^{i\varphi_{\pm 2}}$ 
and $b_{\pm} \rightarrow \sqrt{N_{\pm 1}}\,e^{i\varphi_{\pm 1}}$, where $\{N_s,\varphi_s\}$ 
are conjugate variables, $N_s$ being the number of 
excitons with spin $s=\pm 2$ or $ \pm 1$. The classical Hamiltonian associated with $H_{\rm kin}+H_{\rm int}$ appears as
\begin{eqnarray}\label{eqh1}
{\mathcal H}\!=\epsilon _0 (N_1\!+\! N_{-1})+ v_{dd}(N_2^2 \!\!\!&+&\!\!\! N^2_{-2})\!+\!v_{bb}(N_1^2 + N^2_{-1}) \\
+v_{db}(N_1\!+\!N_{-1}) (N_2 \!+\!N_{-2})  \!\!\!&+& \!\!\! 4g_{db} \sqrt{N_1N_{-1}N_2N_{-2}}\cos \Phi \nonumber
\end{eqnarray}
where $\Phi=\varphi_1+\varphi_{-1}- \varphi_2 -\varphi_{-2}$. 

We then proceed just as above.
We minimize ${\mathcal H}$ at fixed numbers of up and down spin electrons, $N^{(e)}_{\pm 1/2}
=N_{\pm 2}+N_{\mp 1}$ and up and down spin holes, $N^{(h)}_{\pm 3/2}
=N_{\pm 2}+N_{\pm 1}$. For bright excitons photocreated by the absorption of $N$ unpolarized photons, these electron and hole
numbers are equal to $N/2$. So, we end up with $N_1=N_{-1}=N/2-N_2=N/2-N_{-2}$. The system classical energy ${\mathcal E}$
is then given by:
\begin{eqnarray}\label{eqe1}
{\mathcal E}\!=2  N_1\epsilon _0&+& 2v_{dd}(N/2-N_1)^2+2v_{bb}N_1^2  \nonumber \\
&+&4 N_1 (N/2 -N_1)\Big[v_{db}-|g_{db}|\Big] 
\end{eqnarray}

For a fixed exciton number $N$, this quantity is minimum for a number of bright excitons $(+1)$ and $(-1)$ given by  $N^{(0)}_b=N^{(0)}_1+N^{(0)}_{-1}=2N^{(0)}_1$ with 
\begin{eqnarray}\label{eqnb1}
N^{(0)}_b=\frac{N\Big[v_{dd}-v_{db}+|g_{db}|\Big]-\epsilon _0}{\Big[v_{dd}+v_{bb}-2v_{db}+2|g_{db}|\Big]}
\end{eqnarray}
A threshold appears at $N_c=\epsilon _0/[v_{dd}-v_{db}+|g_{db}|]$. 
For an exciton number larger than this threshold, the condensate reads
$(d^{\dag}_{2}d^{\dag}_{-2})^{[N-N^{(0)}_b]/2}(b^{\dag}_{1}b^{\dag}_{-1})^{N^{(0)}_b/2}|vac\rangle$ at zeroth-order in interactions.
However, in the above derivation, we have not handled the problem raised by the degeneracy between $s=\pm 2$ excitons, nor between $s= \pm 1$
excitons. We believe that the treatment of this problem along the arguments of ref.\cite{cbc} should also lead to a linear polarization,
the condensate then reading $([d^{\dag}_{2}+d^{\dag}_{-2}])^{[N-N^{(0)}_b]}
([b^{\dag}_{1}+b^{\dag}_{-1}])^{N^{(0)}_b}|vac\rangle$.

The above treatment of course implies $[v_{dd}-v_{db}+|g_{db}|]$ positive. The effective scatterings
introduced in Eq.(\ref{eq2a}) depend on the "in" and "out" exciton wave functions.
These wave functions are essentially equal for bright and dark excitons; so, $g_{db}$ and the $v$'s can be taken as equal. 
At the Born level, they formally read $\xi (_{0 0}^{0 0})
-\xi^{exch} (_{0 0}^{0 0})$
in terms of the direct and exchange Coulomb scatterings of the composite boson many-body theory \cite{physreport}. For direct Coulomb processes in which the excitons stay in the relative motion ground state,
we have shown that $\xi (_{0 0}^{0 0})=0$, while the exchange Coulomb scattering is given in 3D by $\xi^{exch} (_{0 0}^{0 0})=
-(26 \pi /3)(a_X/L)^3 R_X$, where $R_X$ is the exciton Rydberg, and by $-(8 \pi - 315 \pi^3/512)(a_X/L)^2 R_X$ in 2D \cite{cdb}. As a result, all the scatterings 
considered in Eq.(\ref{eq2a}) are positive.
We thus end up with $[v_{dd}-v_{db}+|g_{db}|] \simeq |g_{db}| $ which indeed is positive.

Using Eq.(\ref{eqnb1}), we then estimate the exciton number threshold as
\begin{eqnarray}\label{eqe1}
N_c \sim \frac{\epsilon _0}{|\xi^{exch} (_{0 0}^{0 0})|} \sim \frac{\epsilon _0}{R_X} \left(\frac{L}{a_X}\right)^2
\end{eqnarray}
This exactly is what can be physically guessed. Indeed, the energy scale for the Coulomb exchange conversion scattering
$g_{db}$ must be the exciton Rydberg $R_X$ while the threshold
is reached when the conversion energy wins over the binding energy $\epsilon _0$.
For $\epsilon _0 \simeq 10 \mu$eV and a carrier mass $0.1$ the free electron mass, Eq.(\ref{eqe1}) gives
a density threshold $N_c/L^2 \simeq 10^9{\rm cm}^{-2}$.



It is worth stressing that, at our estimated value for the exciton number threshold $N_c$, the many-body parameter $\eta$ is smaller than 1. 
This provides a strong justification for our treatment on the dilute side. Even if
the electron-hole plasma is likely to be physically a BCS-like condensate, it seems that the experimental densities will
rather correspond to the intermediate regime $\eta \sim 1$,
analogous to the one found in cold fermionic gases around unitarity. This regime is a very complicated one to deal
with in a quantitative way. An approach from the dilute side, as we have done, could actually be quite appropriate.

In conclusion, we have
proposed a rather unusual BEC-BCS crossover between a dilute and a dense electron-hole system. Indeed, under a density increase, such a system undergoes
a liquid-gas-like first order transition between a dilute exciton gas and a dense electron-hole plasma. At low enough temperature on the dilute side,
a Bose-Einstein condensate of excitons is formed with excitons in a dark state. On the
dense side, the electron-hole plasma undergoes a BCS condensation of excitonic Cooper pairs in a "gray" state,
the condensate having, in addition to its dark component, a coherent bright component which results from Coulomb
exchange scattering between dark and bright excitons. This bright component appears above an exciton density
threshold which falls in the forbidden region of the first order transition. The coherence between these two components should
give rise to an internal Josephson effect with oscillations of the photoluminescence.

 We wish to thank Tony Leggett for very enlightening discussions, during our stay at Urbana, on the formation of Cooper pairs in electron-hole plasma.
 We also are extremely grateful to Michael Stern for sharing with us his experimental results prior to publication, and for
 very inspiring discussions.

\end{document}